\documentclass[]{spie}  

\usepackage{amsmath,amsfonts,amssymb}
\usepackage{graphicx}
\usepackage[colorlinks=true, allcolors=blue]{hyperref}
\usepackage[bottom]{footmisc}
\usepackage{textcomp}

\title{Energy scale calibration and drift correction of the X-IFU}

\author[a]{Edoardo Cucchetti}
\author[b]{Megan E. Eckart}
\author[c]{Philippe Peille}
\author[d]{Cor de Vries}
\author[a]{Fran\c{c}ois Pajot}
\author[a]{Etienne Pointecouteau}
\author[b]{Maurice Leutenegger}
\author[b]{Caroline A. Kilbourne}
\author[b]{Frederick S. Porter}

\affil[a]{\normalsize IRAP, Universit\'{e} de Toulouse, CNRS, CNES, 9 Av. du Colonel Roche, 31400 Toulouse, France}
\affil[b]{NASA/Goddard Space Flight Center, 8800 Greenbelt Rd., Greenbelt, MD 20771, United States}
\affil[c]{CNES, 18 Avenue Edouard Belin 31400 Toulouse, France}
\affil[d]{SRON Netherlands Institute for Space Research, Sorbonnelaan 2, 3584 CA Utrecht, Netherlands}

\authorinfo{Further author information: (Send correspondence to Edoardo Cucchetti)\\ \hspace*{1.6em} Edoardo Cucchetti: E-mail: edoardo.cucchetti@irap.omp.eu}

\begin{document} 
\maketitle

\begin{abstract}
The  \textsl{Athena} X-Ray Integral Field Unit (X-IFU) will provide spatially resolved high-resolution spectroscopy (2.5~eV FWHM up to 7~keV) over the 0.2 to 12~keV energy band. It will comprise an array of 3840 superconducting Transition Edge Sensors (TESs) operated at 90~mK, with an absolute energy scale accuracy of 0.4~eV. Slight changes in the TES operating environment can cause significant variations in its energy response function, which may result in degradation of the detector's energy resolution, and eventually in systematic errors in the absolute energy scale if not properly corrected. These changes will be monitored via an onboard Modulated X-ray Source (MXS) and the energy scale will be corrected accordingly using a multi-parameter interpolation of gain curves obtained during ground calibration. Assuming realistic MXS configurations and using the instrument End-To-End simulator SIXTE, we investigate here both statistical and systematic effects on the X-IFU energy scale, occurring either during ground measurements or in-flight. The corresponding impacts on the energy resolution and means of accounting for these errors are also addressed. We notably demonstrate that a multi-parameter gain correction, using both the pulse-height estimate and the baseline of a pulse, can accurately recover systematic effects on the gain due to realistic changes in TES operating conditions within 0.4~eV. Optimisations of this technique with respect to the MXS line configuration and correction time, as well as to the energy scale parametrization are also show promising results to improve the accuracy of the correction.
\end{abstract}


\section{Introduction}

Scheduled for launch in the beginning of the 2030s on board the future European X-ray observatory \textit{Athena}~[\citen{Nandra2013Athena}], the X-ray Integral Field Unit (X-IFU) [\citen{Barret2016XIFU}] will provide new insights in our understanding of the Hot and Energetic Universe (e.g. galaxy clusters history, Black Hole dynamics or evolution of compact objects). Through its array of $\sim$3840 Transition Edge Sensor (TESs) micro-calorimeters [\citen{Smith2016Pix}] operated at 90~mK, the X-IFU will provide spatially-resolved (5'' over a 5' diameter field-of-view of 5') high-resolution spectroscopy (2.5~eV FWHM energy resolution up to 7~keV) in the soft X-ray band, between 0.2 and 12~keV. 

The detector array of the instrument is composed of X-ray absorbers thermally linked to the TESs. When an incident X-ray photon thermalises in one of these absorbers, it will cause an increase in temperature. As TESs are voltage biased in their superconducting transition, this will cause a rapid change in the overall resistance of the detector and in turn a current pulse. The large number of detectors will be read out using a Frequency Multiplexing scheme (FDM) [\citen{Akamatsu2018FDM}], each pixel's signal being amplitude-modulated on a different carrier. Current pulses obtained after readout retain the information of the incident photon energy and are post-processed in two steps. First, pulses are filtered [\citen{Moseley1988Opt, Bandler2006ResSpace}] to obtain their pulse-height estimate ($PHA$) in detector units. Then, the $PHA$ is transformed into a real energy (in keV) using the so-called energy scale (or gain) function. This function will be calibrated per pixel on the ground using a set of referential lines [\citen{Eckart2018Gain}]. Micro-calorimeters are however extremely sensitive to their environment. Slight changes in their operating parameters may cause significant deviations in their energy scale function. On short time scales, this causes degradations in the energy resolution of the pixel but over longer periods, these changes may result in drifts, eventually causing important systematic effects in the absolute knowledge of the energy scale if not considered. In-flight, such changes will be monitored using the referential lines of the onboard Modulated X-ray Source (MXS) [\citen{deVries2010MXS}] and corrected accordingly during post-processing. After correction, errors on the knowledge of the X-IFU energy scale are required to be lower than 0.4~eV over 0.2~--~7~keV. 

In this contribution, we present several results related to the calibration of the energy scale function of the~X-IFU. Studies are made numerically using the instrument end-to-end simulator SIXTE [\citen{Wilms2014SIXTE}]. In a first part (Sect.~\ref{sec:calib}) the current calibration strategy for the energy scale is introduced and systematic errors made during ground calibration are quantified. Using realistic configurations of the MXS, different techniques to recover changes in the energy scale are then introduced and compared (Sect.~\ref{sec:gain}), to determine the most efficient configuration over a correction time scale of 1~ks. Notably, the effect of the statistics in the referential line(s) and of the choice of the corrected energy bandpass are addressed. Even after an accurate correction, small uncorrected drifts in the energy scale can cause degradations in the energy resolution of the pixels over long periods. The choice of the correction time scale is therefore investigated (Sect.~\ref{sec:res}) to ensure enough statistics in the referential lines used for the correction on the one hand, and to avoid large degradations of the energy resolution on the other. Finally, more natural ways of addressing the energy scale are introduced (Sect.~\ref{sec:real}), as a tentative to include the physical properties of the detectors within the energy scale model, rather than polynomial approaches.

\section{Numerical approach and ground calibration residuals}
\label{sec:calib}

We present in this section our simulation set-up and assess the errors made during ground calibration on the measurements of the energy scale.
\begin{figure}
\centering
\includegraphics[width=0.49\textwidth]{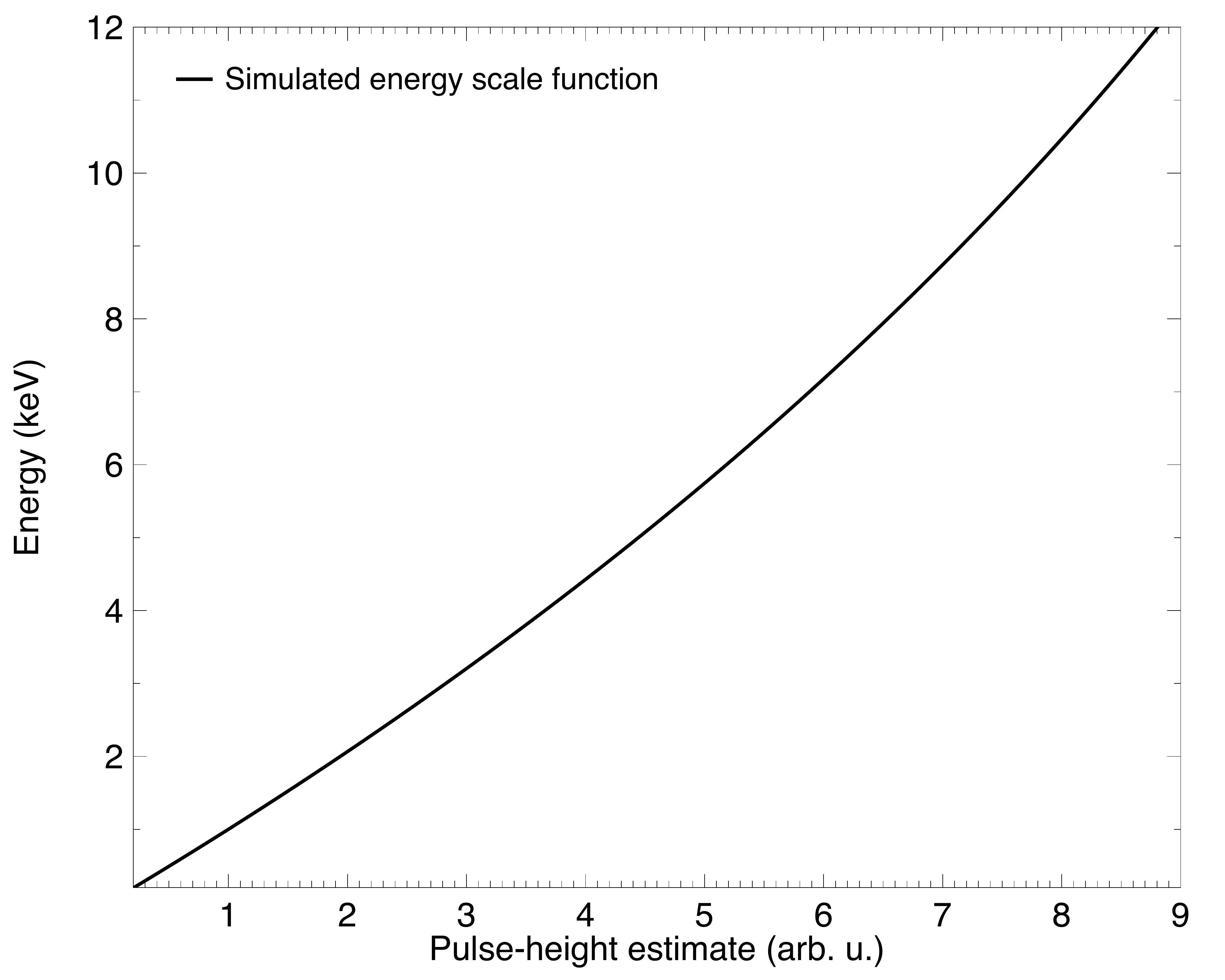}
\includegraphics[width=0.5\textwidth]{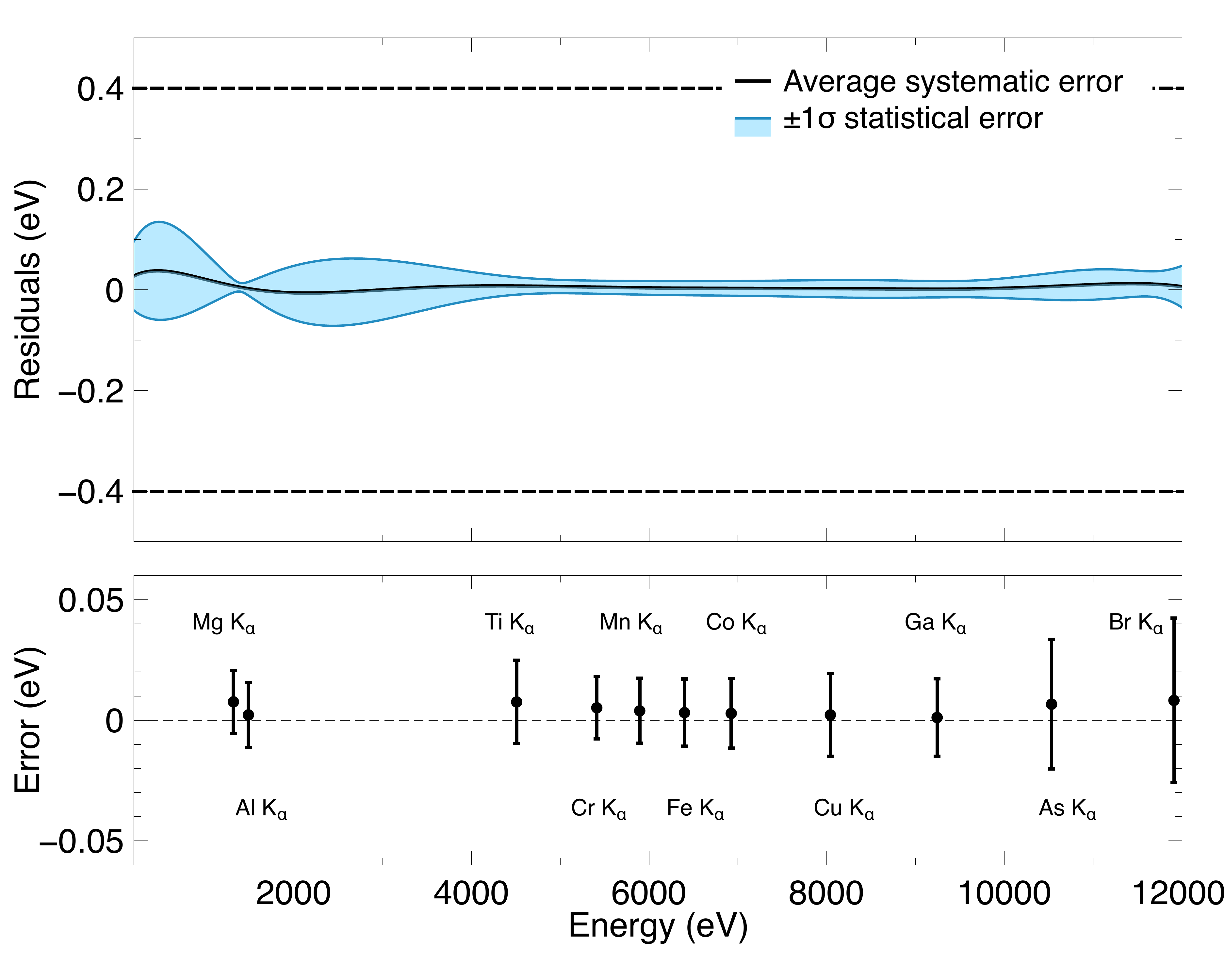}
\caption{(\textit{Left}) Energy scale function simulated using \texttt{tessim} on the SIXTE simulator, for a TES equilibrium set point of $T_{\text{bath}}$=55~mK and  $V_{\text{bias}}$=51.6 nV$_{\text{rms}}$. The function represents the real photon energy (in keV) as a function of the PHA (arbitrary units). (\textit{Right}) Systematic error due to the interpolation of  the energy scale function (in eV) as a function of the energy over the bandpass (eV) and $\pm 1 \sigma$ statistical error for 10 000 counts in each of the calibration line. Dashed lines remind the 0.4~eV requirement on the energy scale. The bottom panel shows the error made on the mean energy of each line. Results are obtained for 1000 iterations of the ground calibration.}
\label{fig:escale}
\end{figure}

\subsection{Simulated energy scale functions}

The energy scale function allows to transform the raw current signal of a filtered pulse into the energy of the incident photon. As no representative data is available for X-IFU detectors, energy scale functions are simulated using the \texttt{tessim} routine [\citen{Wilms2016tessim}] within the instrument end-to-end simulator SIXTE for type LPA2 pixels (pixel type currently baselined for the X-IFU, as presented in [\citen{Smith2016Pix}]). For fixed TES operating conditions, \texttt{tessim} generates the pulse corresponding to a photon of energy $E$ by solving the coupled system of differential equations which regulates the TES (see notably [\citen{Irwin2005TES}] - Equation (3) and (4))\footnote{As systematic effects are investigated here, no noise terms are considered in the differential equations}. Each pulse is then filtered via an optimal filtering template [\citen{Moseley1988Opt}], computed beforehand for 1~keV impact photons. The complete energy scale function can thus be computed by repeating this process over the entire instrumental bandpass (Figure \ref{fig:escale} -- \textit{Left}). Throughout this paper, we assume a TES operating set point with a bath temperature $T_{\text{bath}}$= 55~mK and a bias voltage $V_{\text{bias}}$= 51.6 nV$_{\text{rms}}$. The corresponding energy scale function $G_0$ and its inverse $G_0^{-1}$ are computed via \texttt{tessim} and assumed without errors. Energy scale functions for other operating points can be simulated using the same technique, simply by changing the value of $T_{\text{bath}}$, $V_{\text{bias}}$ or by adding either an excess optical thermal loading power $P_{\text{load}}$ on the TES (e.g. due to infrared or optical loads) or by introducing a linear amplification (multiplicative constant) in the energy scale $L_{\text{amp}}$ (e.g. due to small gain changes in some of the readout chain amplifiers [\citen{Prele2016Gain}]).

\subsection{Ground calibration}

For the X-IFU, the gain function will be calibrated per pixel on the ground for different TES set points. To do so, referential lines at known energies will be measured using the detector array to interpolate the energy scale function. Nevertheless, if the number of counts in the line is low or the energy bandpass sparsely covered in referential lines, the interpolation error could be significant even without any drift. To investigate this effect, we took as starting point the ground calibration of the energy scale made for the soft X-ray spectrometer [\citen{Takahashi2016Hitomi}] on board \textit{Hitomi}. Its energy scale function was derived using a set of referential lines from a calibration rotating target source (RTS) and interpolated using a 4$^{\text{th}}$ degree polynomial, as explained in [\citen{Eckart2018Gain}].  Several of these lines are used here (Table \ref{tab:lines}), assuming a typical count rate of $C = 1$~cts/s/pix over the entire array. 

Each calibration line is simulated in the real energy space for a given exposure time $\Delta t$, and transformed into $PHA$ space using $G_0^{-1}$. The distribution of the line is then fitted in $PHA$ space using log-normal likelihood minimization (C-statistics [\citen{Cash1979}]) leaving as free parameters the heights, widths and centroids of the lines. The results of the fits are then used to interpolate the energy scale function and compare it to $G_0$ generated through \texttt{tessim}. This process is repeated multiple times to determine the statistical uncertainties on the interpolation (Figure \ref{fig:escale} -- \textit{Right}). The highest residual over the energy band for different exposure times is given in Table \ref{tab:resi} for a 7$^{\text{th}}$ degree polynomial, which gave the best results overall. 

\begin{table}[!ht]
\caption{Lines used for the energy scale ground calibration} 
\label{tab:lines}
\begin{center}       
\begin{tabular}{|c|c|} 
\hline
\rule[-1ex]{0pt}{3.5ex}  Line & Profile-weighted energy (keV)  \\
\hline
\rule[-1ex]{0pt}{3.5ex}  Mg K$_{\alpha}$ & 1.32   \\
\hline
\rule[-1ex]{0pt}{3.5ex}  Al K$_{\alpha}$ & 1.49   \\
\hline
\rule[-1ex]{0pt}{3.5ex}  Ti K$_{\alpha}$ & 4.51   \\
\hline
\rule[-1ex]{0pt}{3.5ex}  Cr K$_{\alpha}$ & 5.41   \\
\hline
\rule[-1ex]{0pt}{3.5ex}  Mn K$_{\alpha}$ & 5.89   \\
\hline
\rule[-1ex]{0pt}{3.5ex}  Fe K$_{\alpha}$& 6.40   \\
\hline
\rule[-1ex]{0pt}{3.5ex}  Co K$_{\alpha}$& 6.92   \\
\hline
\rule[-1ex]{0pt}{3.5ex}  Cu K$_{\alpha}$ & 8.04   \\
\hline
\rule[-1ex]{0pt}{3.5ex}  Ga K$_{\alpha}$&  9.24  \\
\hline
\rule[-1ex]{0pt}{3.5ex}  As K$_{\alpha}$&  10.5  \\
\hline
\rule[-1ex]{0pt}{3.5ex}  Br K$_{\alpha}$ &  11.9  \\
\hline
\end{tabular}
\end{center}
\end{table}

As expected, for low number of counts in the lines, the error made on the energy scale is mostly dominated by the statistics of the lines. Even for high exposure times, the error can never be reduced below 0.03~eV over the bandpass due to interpolation errors. The importance of covering the entire bandpass is highlighted when the same exercise is made with no low-energy lines (e.g. simulating the presence of a beryllium valve during calibration phases). In this case (Table \ref{tab:resi}) interpolation errors rapidly climb to $\sim 0.1$~eV when the Mg line is removed, and even above 0.2~eV when all low-energy lines (Mg, Al and Ti) are removed. For the X-IFU, in addition to classical X-ray sources, the use of Electron Beam Ion Traps (EBITs) may be specifically suited to ground calibration, to ensure a wide coverage over the entire energy band with very high spectral resolution lines ($\leq 1$~eV), therefore ensuring the lowest possible residuals during ground calibration.

\begin{table} [tb]
\caption{Maximal ground calibration residual over the energy band (eV) and $\pm 1 \sigma$ statistical error related to the number of counts on the pixel for different line configuration, with either all the lines listed in Table \ref{tab:lines}, without Mg or without the low-energy lines (Mg, Al and Ti) to mimic the effect of a beryllium gate valve during calibration. Results are achieved using 1000 iterations of the simulated ground calibration process} 
\label{tab:resi}
\begin{center}       
\begin{tabular}{|c|c|c|c|} 
\hline
\rule[-1ex]{0pt}{3.5ex}  Counts per pixel & Residuals (all lines - eV) & Residuals (no Mg - eV) & Residuals (no Mg, Al, Ti - eV)  \\
\hline
\rule[-1ex]{0pt}{3.5ex}  5000 & 0.05 $\pm$ 0.12 & 0.10 $\pm$ 0.35 & 0.19 $\pm$ 0.51  \\
\hline
\rule[-1ex]{0pt}{3.5ex}  8000 & 0.04 $\pm$ 0.09 & 0.08 $\pm$ 0.31 &  0.13 $\pm$ 0.46   \\
\hline
\rule[-1ex]{0pt}{3.5ex} 10000 & 0.03 $\pm$ 0.06 & 0.07 $\pm$ 0.27 &  0.12 $\pm$ 0.41   \\
\hline
\rule[-1ex]{0pt}{3.5ex}  12000 & 0.03 $\pm$ 0.05 & 0.07 $\pm$ 0.25 & 0.12 $\pm$ 0.38   \\
\hline
\rule[-1ex]{0pt}{3.5ex}  15000 & 0.03 $\pm$ 0.03 & 0.07 $\pm$ 0.21 & 0.11 $\pm$ 0.33   \\
\hline
\end{tabular}
\end{center}
\end{table}

\section{Energy scale monitoring and correction}
\label{sec:gain}

Slight changes in TES operating condition in-flight may cause variations in the real energy scale function of the detectors. If left uncorrected, these changes may affect the absolute knowledge of the energy scale and therefore create systematic effects on the science. We develop in this section the solutions put in place to monitor and correct such changes during observations and compare different MXS configurations.

\subsection{Monitoring the energy scale}

In-flight monitoring of the energy scale will be performed using a dedicated Modulated X-ray Source (MXS) inspired by the one used on \textit{Hitomi} [\citen{deVries2010MXS}]. The MXS is a high voltage cathode which -- under constant bias -- creates an electron flux towards its anode, where a layer of metal is deposited. With a fine tuning of this layer, the MXS will fluoresce under specific X-ray lines and create stable, referential lines on the detector. The layer deposited at the anode of the MXS can be adapted to create one or multiple lines depending on the calibration needs. To this effect, multiple configurations were simulated using \texttt{GEANT-4} [\citen{Agostinelli2003GEANT}] to ensure their feasibility. The accuracy of the model was compared with real data taken from \textit{Hitomi} soft X-ray spectrometer's MXS, with very consistent results [\citen{DeVries2018MXS}]. Under these assumptions, we use in the rest of the this paper the following four MXS configurations:
\begin{itemize}
\item Single-line MXS (Cu K$_{\alpha}$) with 3 cts/s/pix
\item High-energy MXS (Cr/Cu K$_{\alpha}$) with 1.5 and 2 cts/s/pix respectively
\item Titanium layer MXS (Ti/Cr/Cu K$_{\alpha}$) with  1.5, 1.5 and 2 cts/s/pix respectively
\item Broadband MXS (Si/Ti/Cr/Cu K$_{\alpha}$) with 1, 1.5, 1.5 and 2 cts/s/pix respectively
\end{itemize}

\subsection{Energy scale correction}
\label{sec:corr}

The simplest solution to correct changes in the energy scale is to use the information of a single referential line $E_0$  (e.g., Cu K$_{\alpha}$) at different times (e.g., $t_0$ and $t_1$) and apply a linear stretch over the entire energy band (i.e., multiplicative constant) such that the new energy scale function $G_1$ at time $t_1$ can be written 
\begin{equation}
G_1 = K G_0
\label{eq:1}
\end{equation}
Although accurate for linear drifts, this solution is quickly reaches its limits for nonlinear detectors such as TESs. More advanced techniques (so called ``nonlinear techniques''), were thus introduced to further increase the accuracy of the correction. This approach generalises the concept of Equation~\ref{eq:1} by considering multiple values of the energy scale for different TES set points (e.g., different operating temperatures) and therefore allowing a more accurate interpolation of the new energy scale function (see [\citen{Porter2016Gain}] for a detailed presentation). In the case of DC-coupled detectors such as TESs, the correction can be further improved by using, in addition to the $PHA$ of the lines, the value of the pixel's baseline $B$ (i.e., its set point current in \textmu A) which depends on the operating conditions of the TES  [\citen{Cucchetti2018Gain}]. This allows to extend the nonlinear technique into a two-dimensional space and significantly improve the correction for larger values of changes in the TES operating parameters. 
\begin{figure}
\centering
\includegraphics[width=0.49\textwidth]{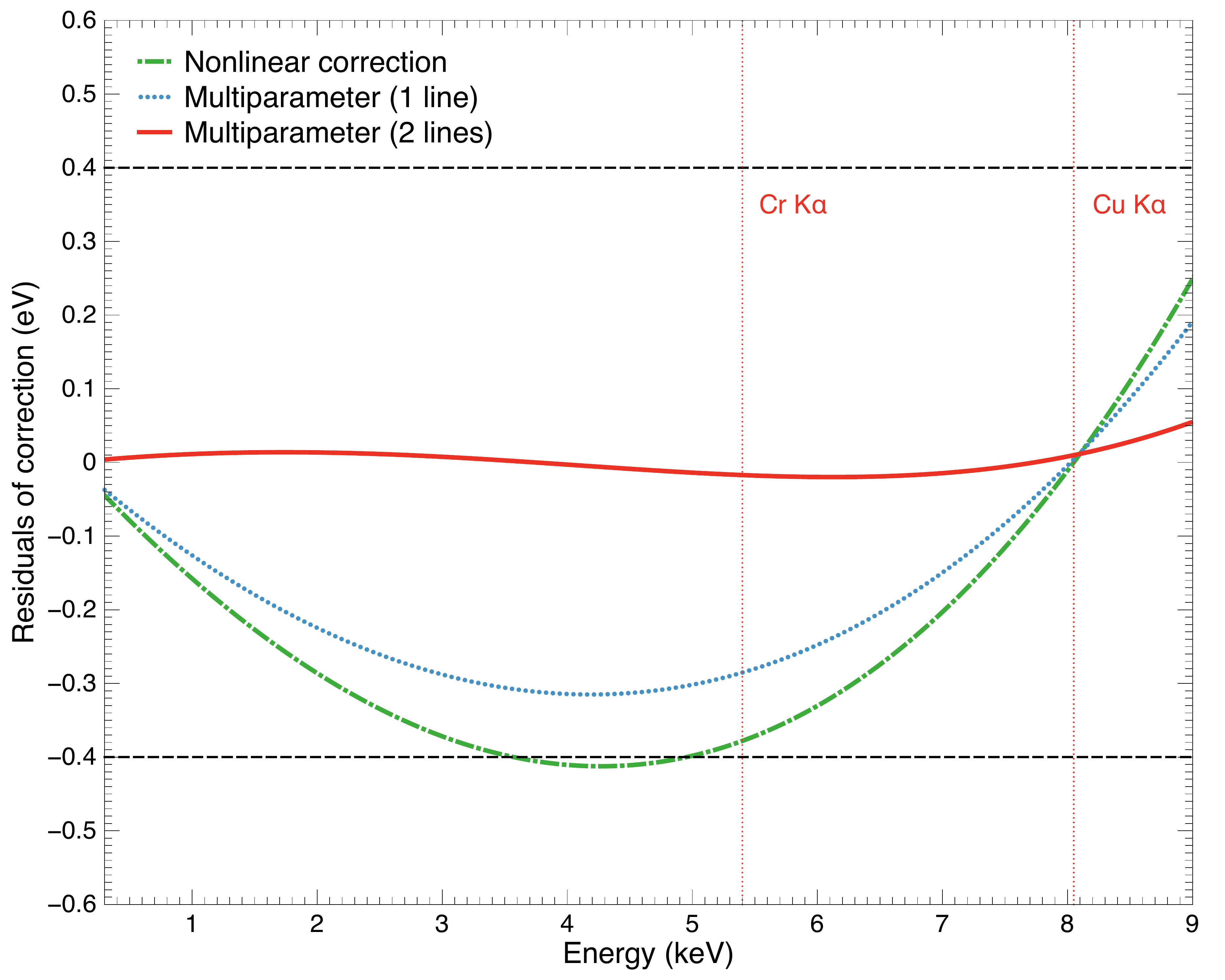}
\includegraphics[width=0.49\textwidth]{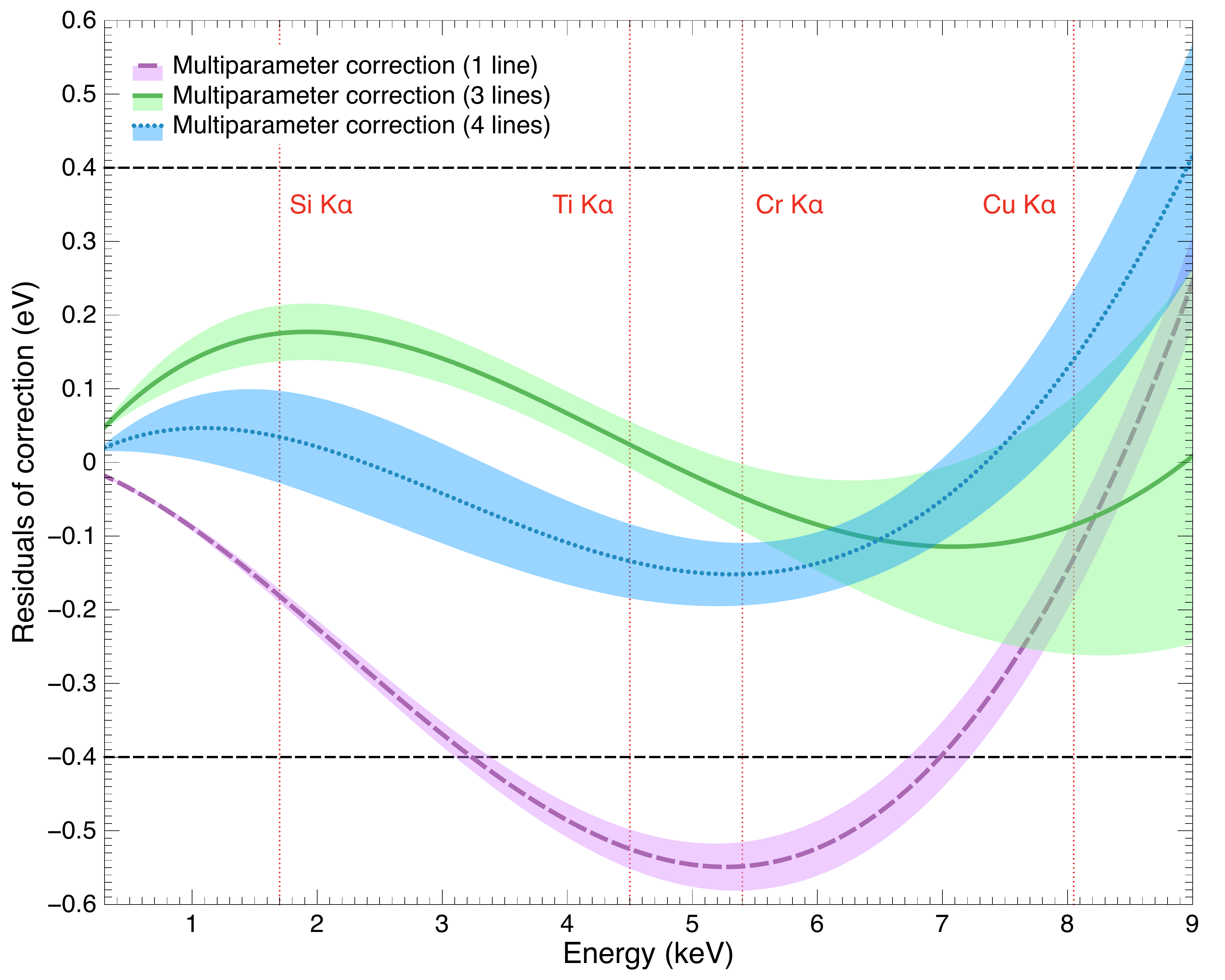}
\caption{(\textit{Left}) Residuals of correction (eV) over the instrumental bandpass (keV) for a 1000 ppm change ($\sim 0.05$~nV$_{\text{rms}}$) in bias voltage ($V_{\text{bias}}$) corrected using the nonlinear correction [\citen{Porter2016Gain}] (green dash-dotted line), the multiparameter correction with one (blue dotted line) or two (red solid line) referential lines. (\textit{Right}) Likewise for a +0.45~pW additional power load ($P_{\text{load}}$) corrected with the multi-parameter method using either one (violet dashed line), three (green solid line) and four lines (dotted blue line) when statistics are included (1~ks correction time, see Sect.~\ref{sec:stats}). The $\pm 1 \sigma$ envelope is filled with the corresponding colors. In both plots, the vertical dotted lines remind the energy of the lines used for the correction, the dashed horizontal lines correspond to the 0.4~eV required for the correction.}
\label{fig:corr}
\end{figure}

In the rest of the paper, we consider this multi-parameter nonlinear technique presented in [\citen{Cucchetti2018Gain}] for the correction, using as `effective' parameters the bath temperature of the TESs, $T_{\text{bath}}$, and a linear multiplication factor, $L_{\text{amp}}$, to maintain a linear degree of freedom in the correction. To apply this correction, we start by simulating six different energy scale functions for different TES set points and interpolate the two-dimensional calibration function $\underline{f_0} : \mathbb{R}^{2} \rightarrow \mathbb{R}^{2}$ at the value of the referential line $E_0$ for which:

\begin{equation}
(PHA_{\text{cal}}(E_0), B_{\text{cal}}(E_0)) = \underline{f_{E_0}} (T_{\text{bath}}, L_{\text{amp}})
\label{eq:2}
\end{equation}
By inverting the function $\underline{f_{E_0}}$ (through a minimisation problem), one can estimate for a given referential energy the value of the effective parameters associated to the change in TES operating conditions, i.e.,

\begin{equation}
\underline{f_{E_0}}^{-1}(PHA_{\text{change}}(E_0), B_{\text{change}}(E_0))) = (T_{\text{bath, change}} (E_0), L_{\text{amp, change}} (E_0))
\label{eq:3}
\end{equation}
these two effective parameters are then used to interpolate over the energy band the corrected value of the energy scale function such that for each value of the energy $E$, the associated $PHA$ is

\begin{equation}
PHA(E) = \underline{f_{E}} (T_{\text{bath, change}} (E_0), L_{\text{amp, change}} (E_0)) \cdot \underline{e_1}
\label{eq:4}
\end{equation}
where $\underline{f_{E}}$ is the function interpolated from the ground calibration energy scale (as done at $E_0$ for Equation~\ref{eq:2}) at the value of energy $E$, taken along its first dimension $\underline{e_1}$. An illustration of the correction algorithm, along with a comparison with the standard nonlinear technique [\citen{Porter2016Gain}] is shown in Figure~\ref{fig:corr} (\textit{Left}). Whenever multiple referential lines are available, this method can be extended by computing energy-dependent parameters (e.g., $ T_{\text{bath, change}} (E), L_{\text{amp, change}} (E)$) in Equation~\ref{eq:4}. This was made by taking a polynomial interpolation of the effective parameters over the energy band using the information of the lines. An example of a correction using multiple lines is also shown Figure~\ref{fig:corr} (\textit{Left}).

\subsection{The effect of statistics}
\label{sec:stats}

The correction of the energy scale will be performed regularly during observations, by integrating the line counts over a given time scale. However, this approach will introduce statistical uncertainties in the overall correction related to the imperfect knowledge of the lines. This effect is included in our simulations using realistic line profiles for each element [\citen{Holzer1997Lines}]: for a given exposure time $\Delta t$, each line is simulated with the same approach developed in Sect.~\ref{sec:calib}, and fitted in $PHA$ space using C-statistics. The profile-weighted $PHA$ of the line is then used as input in the energy scale correction routines presented in  Sect.~\ref{sec:corr}. By performing the correction over the same change in operating conditions a large number of times, the associated statistical error can thus be computed (see the corresponding $\pm 1 \sigma$ error Figure~\ref{fig:corr} -- \textit{Right}).

Using this method, we tested the performance of the correction algorithms assuming a 1~ks correction time for the various MXS configurations. Simulated changes in the TES operating conditions included changes in bath temperature, $T_{\text{bath}}$, voltage bias, $V_{\text{bias}}$, excess power load, $P_{\text{load}}$ or linear drifts of value, $L_{\text{amp}}$. Results for the different correction techniques are shown in Table~\ref{tab:mxs}.
\begin{table} [b]
\caption{Maximal change in TES operating conditions recovered within 0.4~eV (within $\pm 1 \sigma$) over 0.2~--~7~keV and 0.2~--~12~keV using various correction techniques and different MXS configurations for a 1~ks correction time. Parameters considered here are $T_{\text{bath}}$ (in mK), $V_{\text{bias}}$ (in nV$_{\text{rms}}$),  $L_{\text{amp}}$ (in \% of multiplicative factor) and $P_{\text{load}}$ (in pW - always positive). Deviations are given with respect to the TES equilibrium set point, i..e,  $T_{\text{bath}}$=55~mK, $V_{\text{bias}}$=51.6 nV$_{\text{rms}}$,  $L_{\text{amp}}$=0~ppm and  $P_{\text{load}}$=0~pW.} 
\label{tab:mxs}
\begin{center}       
\begin{tabular}{|c|c|c|c|c|c|} 
\cline{2-6}
\multicolumn{1}{c|}{} & \rule[-1ex]{0pt}{3.5ex} Nonlinear correction & \multicolumn{4}{c|}{Multi-parameter nonlinear correction ($T_{\text{bath}}$, $L_{\text{amp}}$)} \\
\hline
\rule[-1ex]{0pt}{3.5ex}  Configuration & Single line [\citen{Porter2016Gain}] & Single-line & High-energy & Titanium layer  & Low-energy \\
\hline
\rule[-1ex]{0pt}{3.5ex}  Line(s) used & Cu & Cu & Cr, Cu & Ti, Cr, Cu & Si, Ti, Cr, Cu \\
\hline 
\multicolumn{6}{c}{}\\
\hline
\rule[-1ex]{0pt}{3.5ex}  Bandpass & \multicolumn{5}{c|}{0.2 -- 7~keV} \\
\hline
\rule[-1ex]{0pt}{3.5ex} $\Delta T_{\text{bath}}$ (mK)  & $\pm$ 2.5 &  $\pm$ 4.0 &  $\pm$ 4.5 &  $\pm$ 5.0 &  $\pm$ 5.5 \\
\hline
\rule[-1ex]{0pt}{3.5ex} $\Delta V_{\text{bias}}$ (nV$_{\text{rms}}$)  & $\pm$ 0.04 &  $\pm$ 0.05 &  $\pm$ 0.4 &  $\pm$ 0.4 &  $\pm$ 0.6 \\
\hline
\rule[-1ex]{0pt}{3.5ex} $\Delta L_{\text{amp}}$ (\%)  & $\pm$ 0.5 &  $\pm$ 1.0 &  $\pm$ 0.7 &  $\pm$ 0.8 &  $\pm$ 1.3 \\
\hline
\rule[-1ex]{0pt}{3.5ex}  $\Delta P_{\text{load}}$ (pW)  & +0.22 & +0.35 & +0.48 & +0.55  & +0.65  \\
\hline 
\multicolumn{6}{c}{}\\
\hline
\rule[-1ex]{0pt}{3.5ex}  Bandpass & \multicolumn{5}{c|}{0.2 -- 12~keV} \\
\hline
\rule[-1ex]{0pt}{3.5ex} $\Delta T_{\text{bath}}$ (mK)  & $\pm$ 2.3 &  $\pm$ 3.5 &  $\pm$ 4.0 &  $\pm$ 4.5 &  $\pm$ 3.7 \\
\hline
\rule[-1ex]{0pt}{3.5ex} $\Delta V_{\text{bias}}$ (nV$_{\text{rms}}$)  & $\pm$ 0.01 &  $\pm$ 0.02 &  $\pm$ 0.07 &  $\pm$ 0.10 &  $\pm$ 0.11  \\
\hline
\rule[-1ex]{0pt}{3.5ex} $\Delta L_{\text{amp}}$ (\%)  & $\pm$ 0.3 & $\pm$ 0.7 & $\pm$ 0.55 &  $\pm$ 0.55 &  $\pm$ 0.45 \\
\hline
\rule[-1ex]{0pt}{3.5ex}  $\Delta P_{\text{load}}$ (pW)  & +0.20 & +0.26 & +0.43 & +0.50  & +0.54  \\
\hline
\end{tabular}
\end{center}
\end{table}

When a finite number of counts of the MXS is included, we notice that the overall accuracy of the correction decreases with respect to systematic-only results found in [\citen{Cucchetti2018Gain}] for all the techniques under consideration.  As expected, the use of multiple lines (notably low-energy lines e..g, Si) generally improves the accuracy of the correction over the energy band, especially for parameters such as the bias voltage which were poorly reconstructed with a single referential line. In fact, variations in TES operating conditions create nonlinear changes in the energy scale over the energy bandpass, which are better described using an energy-dependent effective parameter. However, this is not the case for all parameters, and notably for the multiplicative factor $L_{\text{amp}}$. In this case, we notice that a single line correction with higher statistics ($\sim$ 3000 cts in the Cu K$_{\alpha}$ line) is able to compensate larger changes than a two-line configuration, unlike what is suggested by Figure~\ref{fig:corr} (\textit{Left}), where lines were taken without errors. This can be explained by the larger interpolation error made on the effective parameters when the statistics of the lines are lower. Overall, this error is generally compensated by the information from multiple lines, but may become important for certain parameters which are very sensitive to nonlinear changes (such as the linear multiplicative coefficient  $L_{\text{amp}}$) or in the case of more complex drifts [\citen{Cucchetti2018Gain}]. This error contribution will be at the center of future studies to understand the changes of the parameter over the energy band and therefore reduce this interpolation error. Despite these features, the addition of referential lines certainly helps the broadband recovery of the energy scale. With a better understanding of the nature of the drifts, optimisations of the correction and of the MXS should enable a robust and accurate energy scale correction over 0.2 -- 7~keV.

The additional information introduced by multiple lines comes however at a cost at high energies. As shown also in Table~\ref{tab:mxs}, when a larger correction band is considered (i.e. 0.2 -- 12~keV instead of 0.2 -- 7~keV) the absence of high-energy referential lines and the simplistic polynomial interpolation of the parameters reduces the efficiency of the correction (either with or without statistics included) as errors tend to be high for the extrapolated parts of the energy scale (see also blue and green curves Figure~\ref{fig:corr} -- \textit{Right}). Although such an accurate correction may not be required for the upper part of the bandpass, the final choice of the MXS configuration shall be pondered by the scientific needs driving the absolute knowledge of the energy scale over the high-energy band.

\section{Degradation of the energy resolution}
\label{sec:res}
\begin{figure}[b]
\centering
\includegraphics[width=0.49\textwidth, clip=True, trim={1cm 2cm 3cm 1cm}]{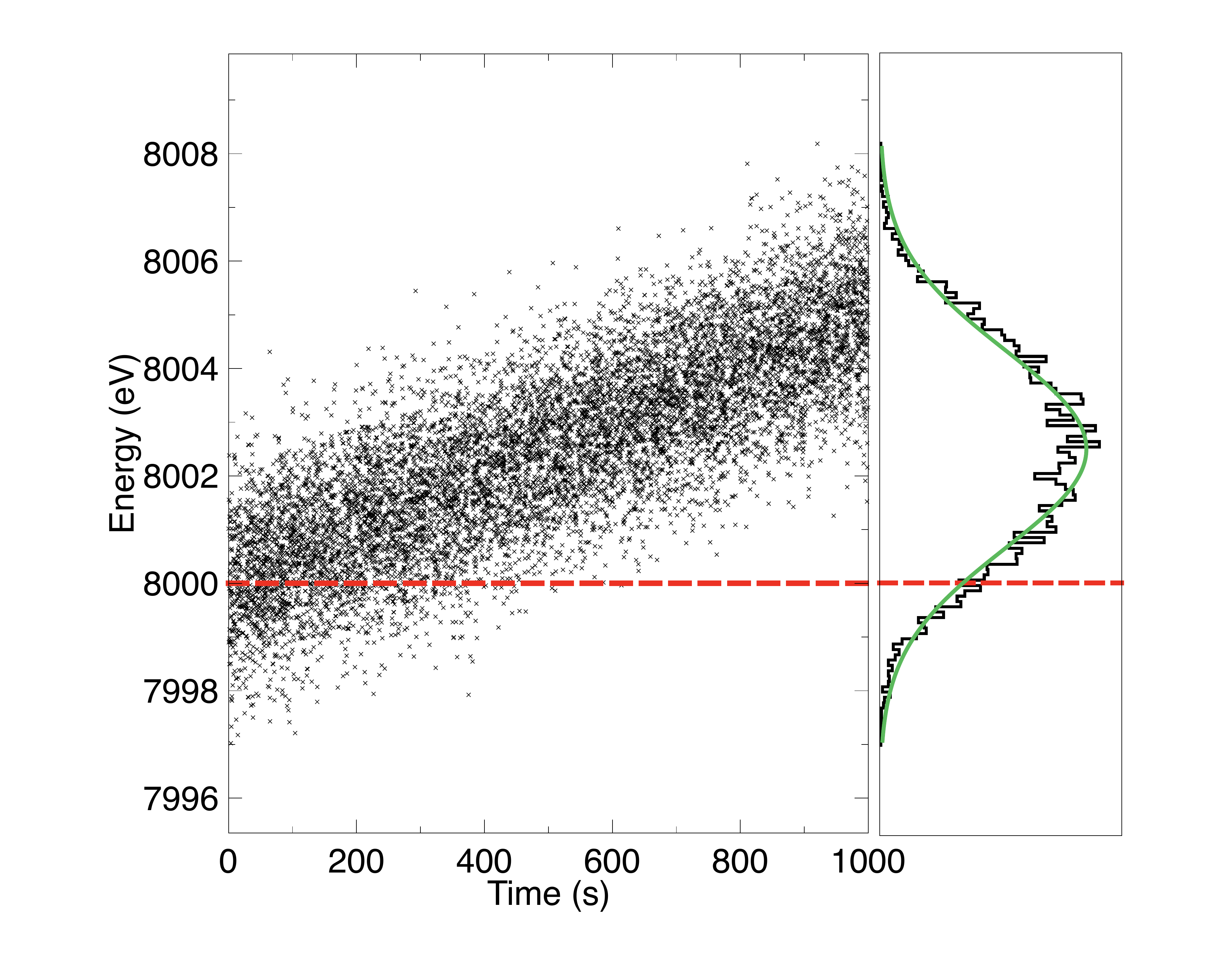}
\includegraphics[width=0.49\textwidth]{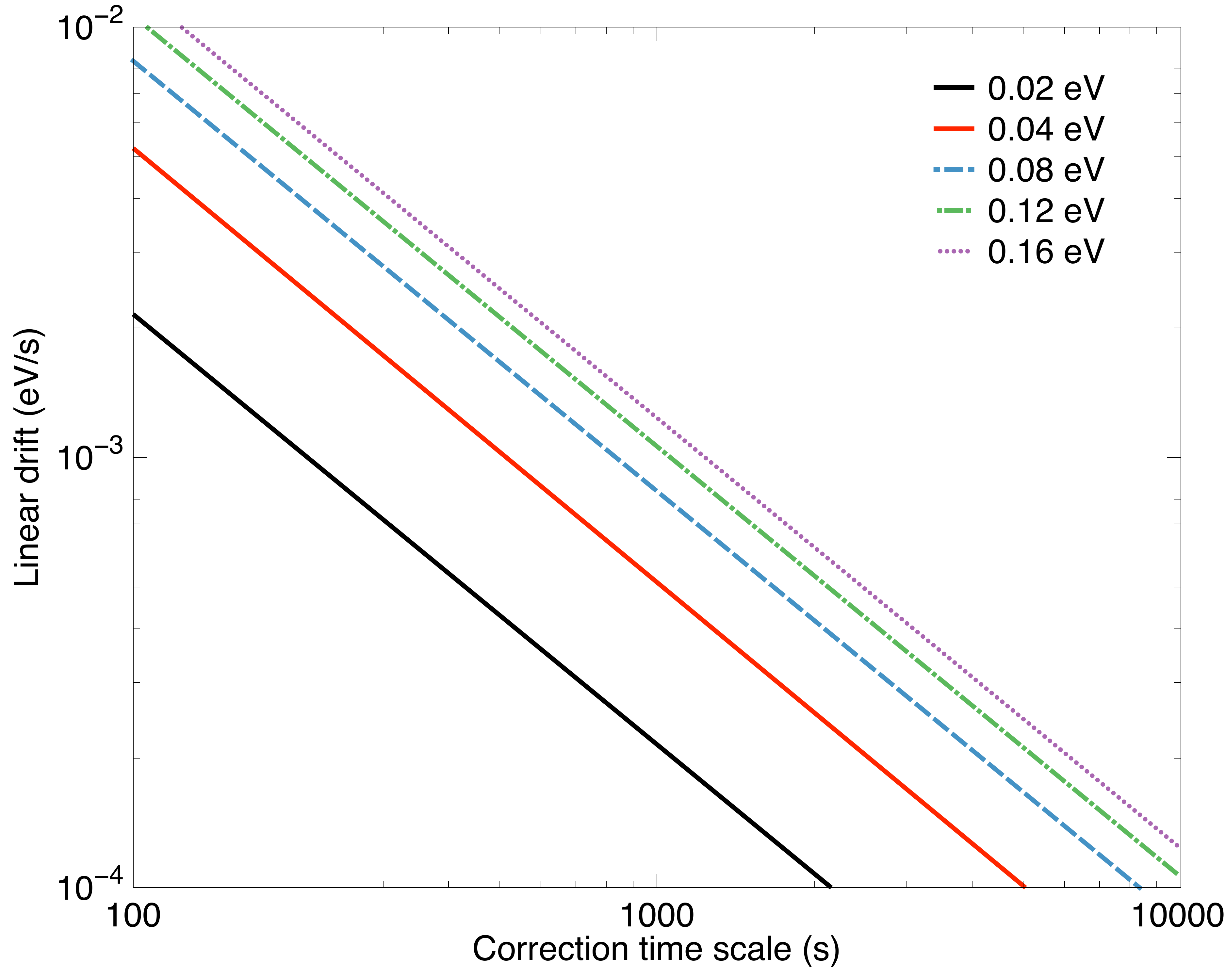}
\caption{(\textit{Left}) Illustration of a residual linear degradation of a 2.5~eV (FWHM) wide Gaussian line centred at $ E_0=8$~keV (dashed red line). The drift value is purposely exaggerated in the plot for clarity, but is taken below $4 \times 10^{-4}$~eV/s in the study (for a 1~ks correction time). The left panel shows the energy of the events across time, the right panel their distribution (shifted and broadened Gaussian). The green solid line indicates the Gaussian best fit of the line. (\textit{Right}) Contours of the energy resolution degradation as a function of the correction time scale and the linear drift. Below (resp. above) these curves the degradation is lower (resp. higher) than the indicated value.}
\label{fig:deg}
\end{figure}

Despite theoretical improvements in the correction methods, a perfect correction of the energy scale over time remains ideal. Even after the correction, small residual drifts may remain. Although this error should be low on the energy scale (required below 0.4~eV over the bandpass), it may cause degradations of the detector's energy resolution (see Figure~\ref{fig:deg} -- \textit{Left} for illustration purposes) whenever the residual uncorrected drift is important. This effect is related to the time scale of the energy scale correction: if the correction is too regular (i.e., short correction time scale), the referential lines used for the correction (see Sect.~\ref{sec:stats}) will not have enough counts to ensure a statistically accurate correction over the energy bandpass. On  the other hand, if it is too sparse (i.e., long correction time scales), the energy scale correction will average the effects of the drifts over time, possibly leaving important short time scale drifts which could degrade the energy resolution. 

To investigate the effect of the correction time, we simulated a Gaussian line (2.5~eV FWHM) of energy $E_0$ to which we apply, as a first approach, a constant linear drift (in eV/s) to simulate a residual drift after energy scale correction. The resulting line is then fitted using a Gaussian model. Figure~\ref{fig:deg} (\textit{Right}) shows the absolute degradations of the energy resolution of the line as a function of the residual drift and the correction time. Under these assumptions, to remain within a 0.02~eV absolute degradation of the energy resolution over the correction time scale, drifts should be lower than $2 \times 10^{-4}$~eV/s if a correction time of 1~ks is assumed. If this condition is not met, although the energy scale correction should ensure errors below 0.4~eV (i.e., residual linear drifts will be below $4 \times 10^{-4}$~eV/s for 1~ks),  important degradations of the energy resolution may occur. Further, depending on the nature of the drifts and the corresponding correction, the new value of the energy scale may also present sharp differences with the real value (see e.g. the derivative of the residuals Figure~\ref{fig:corr} -- \textit{Right}). Locally, this may create potentially important effects on the energy resolution over the time scale of the correction. 

Though preliminary, these results show that despite an accurate correction of the energy scale, effects on the energy resolution may be significant. The time scale of the correction is thus an interesting parameter to leverage to decrease this effect, but also to improve the overall energy scale correction. In practice, the time scale of the correction will depend on the choice of the MXS (lines, count rate) and on the efficiency of the correction algorithm. The actual energy resolution degradation related to uncorrected drifts will be very difficult to determine in-flight. An in-depth analysis of the drift residuals (particularly their time dependence which was simplistically considered linear here) is thus required to understand, model and mitigate these effects.

\section{Towards a more realistic energy scale description?}
\label{sec:real}

The overall interpolation of the energy scale function is currently performed using polynomials. Although the residuals of this interpolation are very accurate ($\lesssim 10^{-3}$~eV for Figure~\ref{fig:escale} -- \textit{Left}), they do not take into account the physical properties of the detectors. We investigate in this section ways to provide a more natural description of the energy scale function. 

In the case of optimal filtering [\citen{Moseley1988Opt}], the mathematical relationship between the pulse-height $PHA$ and the energy $E$ is derived from the filtered pulse. For a pulse without noise, the link corresponds to the dot product of the pulse shape at energy $E$ as a function of time $t$ -- noted $P(t,E)$ -- with the pulse sample -- $S(t)$ (in our case the pulse for 1~keV) -- normalised to the integral of the pulse sample (assumed to be 1 here without any loss of generality). Thus the inverse energy scale function is given by 
\begin{align}
PHA(E) &= \left\langle P(t,E) \cdot S(t) \right\rangle = \int_{0}^{\infty}  P(t,E) \cdot S(t)dt
\label{eq:5}
\end{align}

\begin{figure}
\centering
\includegraphics[width=0.49\textwidth]{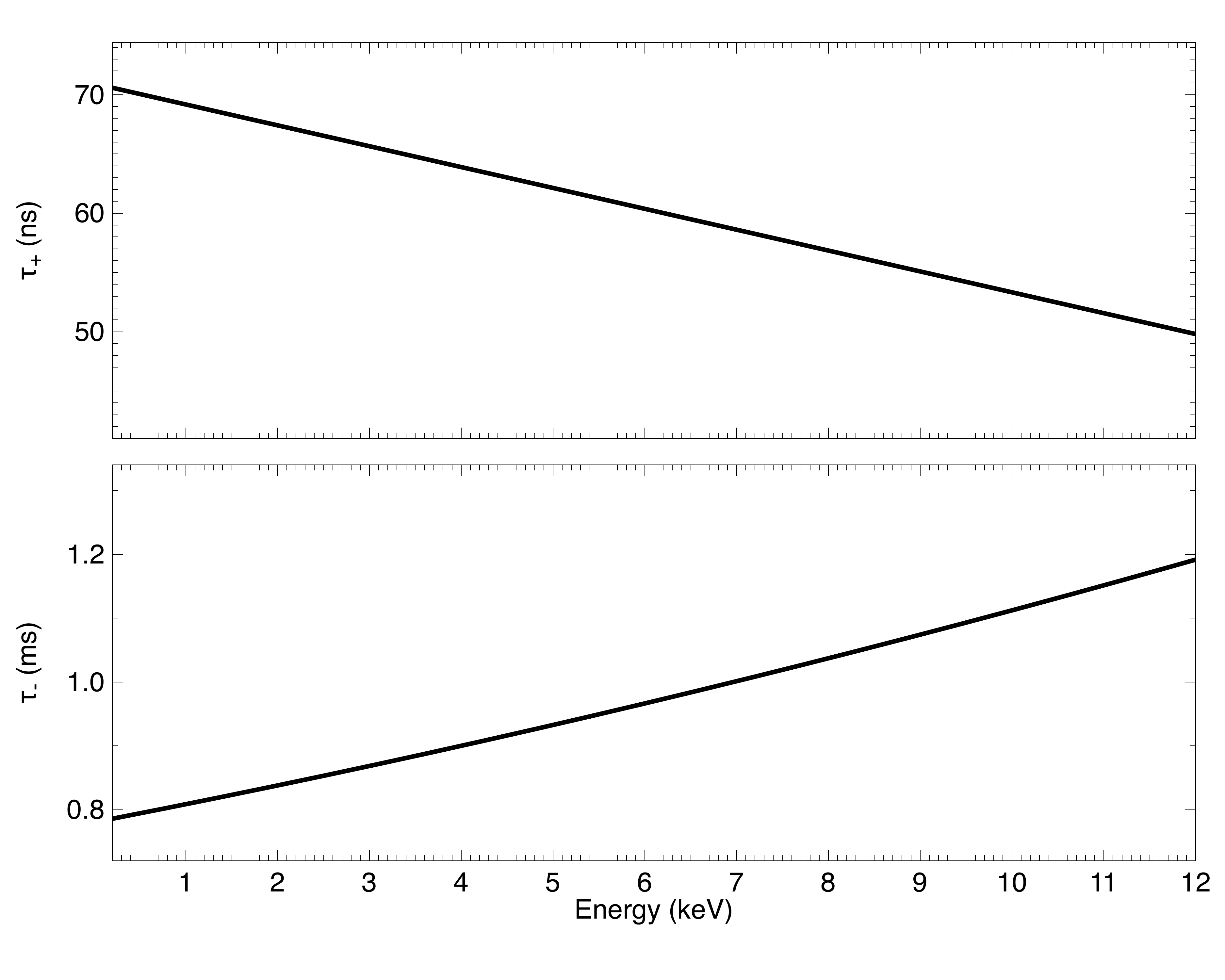}
\includegraphics[width=0.49\textwidth, clip=True]{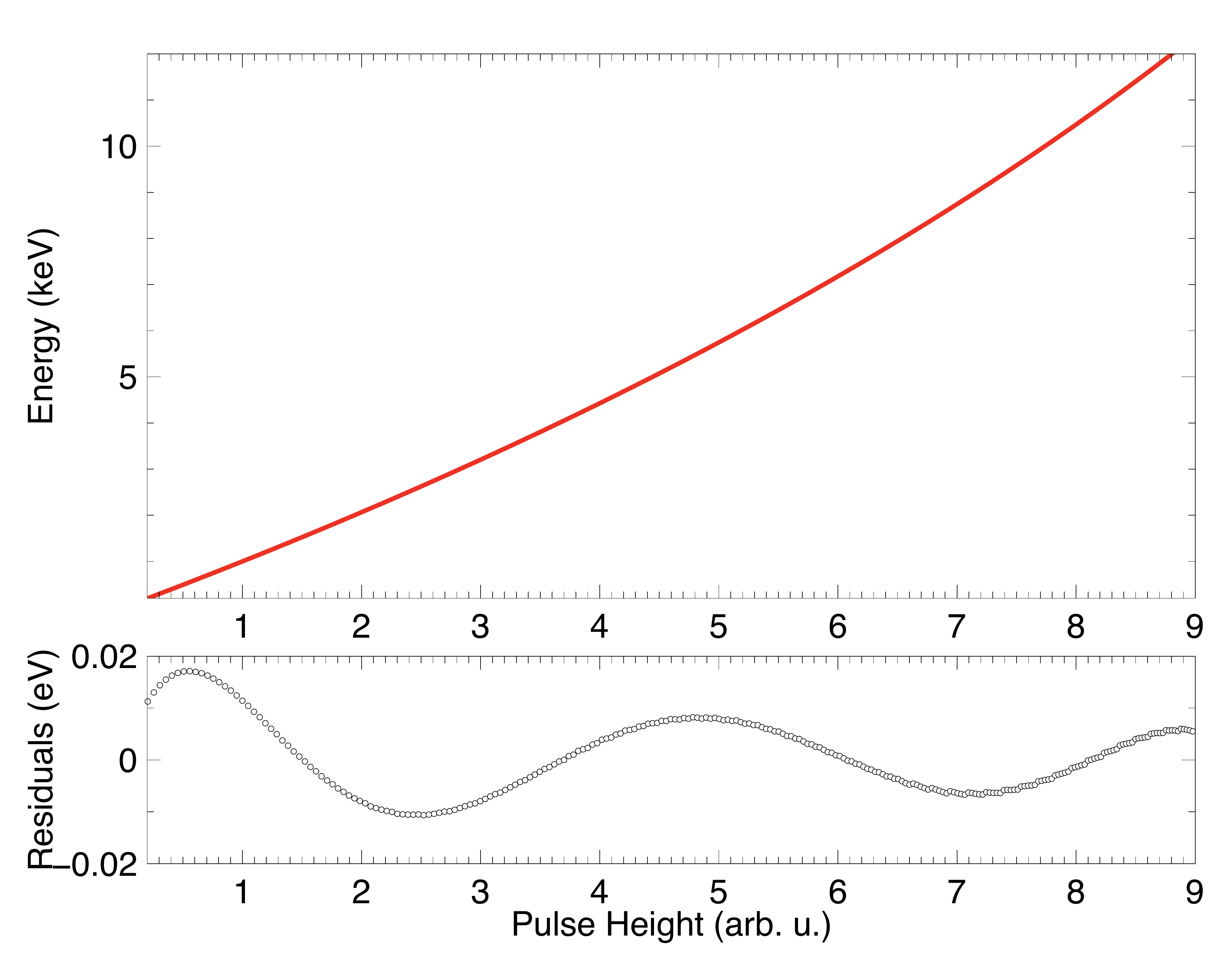}
\caption{(\textit{Left}) Rise time $\tau_{+}$ (ns -- \textit{Top}) and fall time $\tau_{-}$ (ms -- \textit{Bottom}) of the current pulse as a function of the energy of the incident photon obtained using the end-to-end simulator. (\textit{Right}) Energy scale function interpolated using Equation~\ref{eq:7} along with the residuals over the energy bandpass with respect to the real energy scale function shown in Figure~\ref{fig:escale} (\textit{Left}).}
\label{fig:nat}
\end{figure}

Ideally, if the expression of the pulses $P(t,E)$ could be determined analytically, the function $PHA(E)$ could also be derived. As TESs obey a system of coupled differential equation with no apparent analytical solution [\citen{Irwin2005TES}], estimates of the current pulse can only be obtained after linearisation such that:
\begin{align}
P(t,E) \propto (e^{-t/\tau_+} - e^{-t/\tau_-}) E
\label{eq:6}
\end{align}
where $\tau_{+}$ and $\tau_{-}$ are respectively the rise and fall time of the current pulses. As demonstrated in [\citen{Peille2016Pulse}], simulated pulse shapes are very similar once normalised, with a difference visible on the fall time and rise time of the pulses.  In fact, these times are related to the energy of  the incident photon's energy (Figure~\ref{fig:nat} -- \textit{Left}), both of which are very well represented by second order polynomials. Using this information and the expression of the pulse shape, we find after integration in Equation~\ref{eq:5}
 \begin{equation}
 \small
PHA(E) \propto E \left[ \tau_+(E) \left( \frac{\tau_+(E_0)}{\tau_+(E)+\tau_+(E_0)} -  \frac{\tau_-(E_0)}{\tau_+(E)+\tau_-(E_0)} \right) \\ 
+ \tau_-(E) \left( \frac{\tau_-(E_0)}{\tau_-(E)+\tau_-(E_0)} -  \frac{\tau_+(E_0)}{\tau_-(E)+\tau_+(E_0)} \right) \right]
\label{eq:7}
\end{equation}
where $E_0$ corresponds to the reference energy used for the optimal filtering (here E$_0$=1~keV). By fitting the energy scale function found using \texttt{tessim} (Figure~\ref{fig:escale} -- \textit{Left}) we find that the expression shown in Equation~\ref{eq:7} provides accurate results, with residuals of the order of $10^{-2}$~eV (Figure~\ref{fig:nat} -- \textit{Right}). Although less accurate than the polynomial counterpart, this approach takes into account a more physical description of the energy scale function. Deviations from the real shape are likely related to the heuristic approach used here, which takes as starting point the linearised solution and assumes a good knowledge of the characteristic times of the detectors.  These results should now be tested and integrated in the previous algorithms of energy scale correction. 

\section{Conclusion}

The accurate knowledge of the energy scale function, linking the shape of the current pulses seen by TESs and the energy of their incident photon, will be crucial for the success of an instrument such as the X-IFU. Starting from numerical simulations carried out using the end-to-end simulator of the instrument SIXTE, we investigated the calibration and correction of the energy scale function for the X-IFU, accounting for both statistical and systematic effects. Notably, we demonstrated that using a multi-parameter nonlinear technique, the energy scale function can be corrected within the required 0.4~eV even for large drifts in the operating parameters (see Sect.~\ref{sec:stats}). Further improvements in the correction may be achieved by optimising the MXS referential lines and the correction time, which also drives the effective energy resolution of the TESs. Finally, a more physical approach of the energy scale was introduced. Though preliminary, the perspective of a more natural expression of the energy scale would allow a significant decrease in the calibration residuals on the ground and strong improvements in the energy scale correction, both of which are worth investigating. All these results were obtained numerically and should now be verified and improved using measurements from representative TESs.

\bibliography{Gain_paper.bib} 
\bibliographystyle{spiebib} 

\end{document}